\documentclass[12pt,preprint]{aastex}

\def\lsim{\mathrel{\rlap{\lower4pt\hbox{\hskip1pt$\sim$}}
    \raise1pt\hbox{$<$}}}                
\def\gsim{\mathrel{\rlap{\lower4pt\hbox{\hskip1pt$\sim$}}
    \raise1pt\hbox{$>$}}}                

\newcommand{\ergs}{{\mathrm{erg~s^{-1}}}}

\newcommand{\lya}{Ly$\alpha$}

\newcommand{\degsq}{{\mathrm{deg^{2}}}}

\usepackage{lscape}
\usepackage{graphicx}
\usepackage{amsmath}

\usepackage{natbib, epsfig}

\shorttitle{QLF at $z\sim 4$}
\begin{document}


\title{The Faint End of the Quasar Luminosity Function at $z\sim 4$: Implications for Ionization of the Intergalactic Medium and Cosmic Downsizing\altaffilmark{1} }

\author{Eilat Glikman\altaffilmark{3,4}, S.~G. Djorgovski\altaffilmark{2}, Daniel Stern\altaffilmark{5}, Arjun Dey\altaffilmark{6}, Buell T. Jannuzi\altaffilmark{6}, Kyoung-Soo Lee\altaffilmark{3,7}}

\altaffiltext{1}{The data presented herein were obtained at the W.M. Keck
Observatory, which is operated as a scientific partnership among the
California Institute of Technology, the University of California and the
National Aeronautics and Space Administration. The Observatory was made
possible by the generous financial support of the W.M. Keck Foundation.}

\altaffiltext{2}{Astronomy Department, California Institute of Technology,
Pasadena, CA 91125}

\altaffiltext{3}{Department of Physics and Yale Center for Astronomy and Astrophysics, Yale University, P.O. Box 208121, New Haven, CT 06520-8121; email: eilat.glikman@yale.edu}

\altaffiltext{4}{NSF Astronomy and Astrophysics Postdoctoral Fellow}

\altaffiltext{5}{Jet Propulsion Laboratory, California Institute of Technology, Pasadena, CA 91109}

\altaffiltext{6}{National Optical Astronomy Observatory, 950 N. Cherry Ave., Tucson, AZ 85719}

\altaffiltext{7}{Jaylee and Gilbert Mead Fellow}

\begin{abstract}
We present an updated determination of the $z\sim4$ QSO luminosity function (QLF), improving the quality of the determination of the faint end of the QLF presented in \citet{Glikman10}.  We have observed an additional 43 candidates from our survey sample, yielding one additional QSO at $z=4.23$ and increasing the completeness of our spectroscopic follow-up to 48\% for candidates brighter than $R=24$ over our survey area of 3.76 $\degsq$. 
We study the effect of using $K$-corrections to compute the rest-frame absolute magnitude at 1450\AA\ compared with measuring $M_{1450}$ directly from the object spectra.  We find a luminosity-dependent bias: template-based $K$-corrections overestimate the luminosity of low-luminosity QSOs, likely due to their reliance on templates derived from higher luminosity QSOs.  Combining our sample with bright quasars from the Sloan Digital Sky Survey and using spectrum-based $M_{1450}$ for all the quasars, we fit a double-power-law to the binned QLF. Our best fit has a bright-end slope, $\alpha = Ð3.3\pm0.2$, and faint-end slope, $\beta = Ð1.6^{+0.8}_{-0.6}$.  Our new data revise the faint-end slope of the QLF down to flatter values similar to those measured at $z\sim3$.   The break luminosity, though poorly constrained, is at $M^* = -24.1^{0.7}_{-1.9}$, approximately $1-1.5$ mag fainter than at $z\sim 3$.  This QLF implies that QSOs account for about half the radiation needed to ionize the IGM at these redshifts.
\end{abstract}

\defcitealias{Glikman10}{Paper~I}

\section{Introduction}
The primary observable that traces the evolution of QSO populations is the QSO luminosity function (QLF) as a function of redshift. The QLF can be well represented by a ÒbrokenÓ power-law: $\Phi(L) = \phi^* [(L/L^*)^\alpha + (L/L^*)^\beta]^{-1}$ , where $L^*$ is the ÒbreakÓ luminosity.  Current measurements poorly constrain the corresponding break absolute magnitude to be  $M^*_{1450} \simeq -25$ to $-26$ mag, at $\lambda_{\rm rest} = 1450$\AA.  The bright-end slope, $\alpha$, appears to evolve and flatten toward high redshifts, beyond $z\sim 2.5$ \citep{Richards06}.   The faint-end slope, $\beta$,  has been measured to be around $-1.7$ at $z\lesssim 2.1$; it is poorly constrained at higher redshifts, but appears to flatten at $z\sim 3$ \citep{Hunt04,Siana08}.  At yet higher redshifts, the situation is much less clear due to the relatively shallow flux limits of most surveys to date \citep[e.g.,][]{Wolf03, Richards06}. The shape of the QLF at $z > 3$ is still not well measured, and the evolution of $L^*$ and the faint-end slope remain poorly constrained.

We recently measured the faint end of the QLF at $z \sim 4$ and found an unexpected result: the QLF appeared to rise steeply towards faint luminosities, steeper than at lower redshifts, and in apparent conflict with some theoretical models \citep[hereafter, Paper~I]{Glikman10}. This implied a more complex co-evolution of AGN and galaxies in the first 2 Gyr of the Universe, and suggested that AGN are a more significant contributor to the metagalactic ionizing UV flux than previously estimated.  However, the faint-end slope computed in \citetalias{Glikman10} relied on the faintest bin, which was highly incomplete with only two QSOs confirmed. 
Recently, \citet{Ikeda10} measured the faint end of the QLF from $z\sim 4$ QSOs in the COSMOS field and found a lower space density by a factor of $\sim$ 3 to 4 and a flatter slope.  The COSMOS survey covers 43\% smaller area than our survey and has three times fewer QSOs (8 versus 24).   In this paper we present new spectroscopic observations from our survey which increase our completeness and yield an improved estimate of the $z\sim 4$ QLF down to $M_{1450}=-21$ mag, primarily due to our improved determination of the contamination rates
in the lowest luminosity bin used to compute the QLF.

We use standard cosmological parameters throughout the paper: $H_0=70$ km s$^{-1}$ Mpc$^{-1}$, $\Omega_M=0.30$, and $\Omega_\Lambda=0.70$.

\section{Sample Selection and Previous Results}

We directly measured the $z=4$ QLF by taking advantage of two publicly available, deep, multiwavelength imaging surveys.  The selection methodology is described in detail in \citetalias{Glikman10}.  Our QSO sample is derived from the NOAO Deep Wide-Field Survey \citep[NDWFS;][]{Jannuzi99} and the Deep Lens Survey \citep[DLS;][]{Wittman02}, which reach magnitude limits of $R\gtrsim24$ and provide multiwavelength imaging in $B$, $V$, $R$, $I$, and $z$, depending on the survey.  We selected subfields from these surveys amounting to a total area of 3.76 $\degsq$ (1.71 $\degsq$ in NDWFS and 2.05 $\degsq$ in DLS).  We created source catalogs from these images using SExtractor \citep{Bertin96} in dual-image mode, requiring a detection in the $R$-band.

To derive and understand our selection function, we built a library of model QSO spectra spanning the redshift range $3.5 < z < 5.2$, including Monte Carlo realizations of \lya\ forest absorbers along the line-of-sight, continuum  slope, and \lya\ equivalent width, to determine the location of these QSOs in color-color space.  We applied color cuts to our catalogs, resulting in 148 QSO candidates, 30 of which are brighter than $R=23$.  In \citetalias{Glikman10} we reported 28 spectra, concentrating on the brightest sources, of which 23 were QSOs with $3.74 < z < 5.06$.  Twenty of these are brighter than $R=23$.  Interestingly, two QSOs at $z\sim 4$ had anomalous \ion{N}{4}] 1486\AA\ emission, suggesting the presence of very massive, metal-poor star formation in their hosts \citep[i.e., Population III;][]{Glikman07}.

We used the modeled QSO spectra to compute the selection probability of a QSO in our survey as a function of $R$ magnitude and redshift. This allowed us to compute the volume density of quasars as a function of luminosity, $\Phi(M_{1450},z =4)$, using the $1/V_a$ method \citep{Page00}.  


We computed the QLF at a fiducial wavelength that, for high-redshift, optically-selected QSOs, is 1450\AA. This wavelength lies in a part of the rest-frame ultraviolet that is free from emission lines, and is a proxy for the continuum luminosity.  Lower redshift surveys typically derive the QLF at rest-frame $B$-band \citep[e.g.,][]{Pei95,Croom04} while \citet{Richards06} use the absolute magnitude of a $z=2$ quasar in the $i$-band, $M_i(z=2)$, for the SDSS QLF.  

A standard way to convert an observed magnitude to a monochromatic rest-frame absolute magnitude at 1450 \AA\ is to add an appropriate $K$-correction, which is often computed either from a QSO template or simulated QSOs, such as our library of model QSO spectra \citepalias[see][]{Glikman10}.
At low redshifts, optical spectra do not cover rest-frame 1450 \AA\ and $K$-corrections are the only option.  However, if rest-frame 1450 \AA\ is covered by the spectrum of a quasar, a more direct way to compute $M_{1450}$ is to convolve a spectrum that has been shifted to the rest frame with a top hat filter centered at 1450 \AA, and apply a distance modulus.  To account for slit losses, we convolve our spectra with a photometric filter ($I$ or $z$, depending on the survey) and scale the fluxes to the image-based photometry.  In \citetalias{Glikman10} we measured $M_{1450}$ for our QSOs using both methods: directly from the spectra and by applying a $K$-correction determined from our simulated model spectra. We found that $M_{1450}$ measured directly from the spectra were $\sim 0.3$ magnitudes fainter, on average, than $M_{1450}$ computed with $K$-corrections.  


In \citetalias{Glikman10} our spectroscopic completeness for $R<23$ was 73\% and we computed the QLF using only these brightest QSOs.  We found that regardless of the method for computing $M_{1450}$, the faint-end slope of the QLF was steeper than at lower redshifts when fit by a single power-law, $\Phi(M)=\Phi^* 10^{-0.4(\beta-1)M}$.  The best-fit slopes were $\beta=-1.6\pm0.2$ and $\beta=-1.2\pm0.2$ for the $K$-correction-based and spectrum-based $M_{1450}$, respectively.  Once we added the three QSOs with $R>23$ (whose spectroscopic completeness was only 5\%) the QLF became even steeper with $\beta = -2.0 \pm 0.2$ and $\beta=-2.5\pm0.2$, for the $K$-correction-based and spectrum-based $M_{1450}$, respectively.  

These steep slopes persisted when we combined the faint QSOs with the QLF derived from SDSS quasars in \citet{Richards06} and \citet{Fontanot07}, and a double power-law fit to the combination of the bright-end QLF from \citet{Richards06} and the faint-end points based on spectrum-based $M_{1450}$ yielded a bright-end slope, $\alpha = -2.4 \pm 0.2$, and faint-end slope, $\beta = -2.3 \pm 0.2$, without a well-constrained break luminosity; this was effectively a single power law.  

There were two immediate cosmological implications to this result: (1) models of the evolution of faint AGN at $\sim 1$ Gyr after the end of the reionization, and possibly all the way into the reionization era, would need to be revised; (2) AGN were a more significant contributor to the metagalactic ionizing UV flux at these epochs, affecting the evolution of the IGM \citep{Haiman01,Wyithe03a,Shankar07}.  However, as emphasized in \citetalias{Glikman10}, these results were based on only a few QSOs in the faintest bin, $23<R<24$.

\section{Additional Spectroscopy}

We sought to improve the measurement of the faint-end slope by obtaining additional spectra of $R>23$ candidates.  On UT 2010 May 17-18 we obtained 38 additional spectra with LRIS on Keck I.  To maximize our efficiency, we observed using slitmasks, selecting candidates that lay within a single slit-mask ($4\arcmin \times 8\arcmin$).  We observed 22 candidates in NDWFS, 19 of which were fainter than $R=23$ magnitudes.  We observed 16 DLS candidates, all fainter than $R=23$.  In addition, we obtained 5 spectra with DEIMOS on UT 2010 May 14 of NDWFS candidates as part of an observing program studying $z\sim 4$ LBGs in the Bo\"{o}tes field \citep{Lee10}.

\begin{figure}
\epsscale{0.8}
\plotone{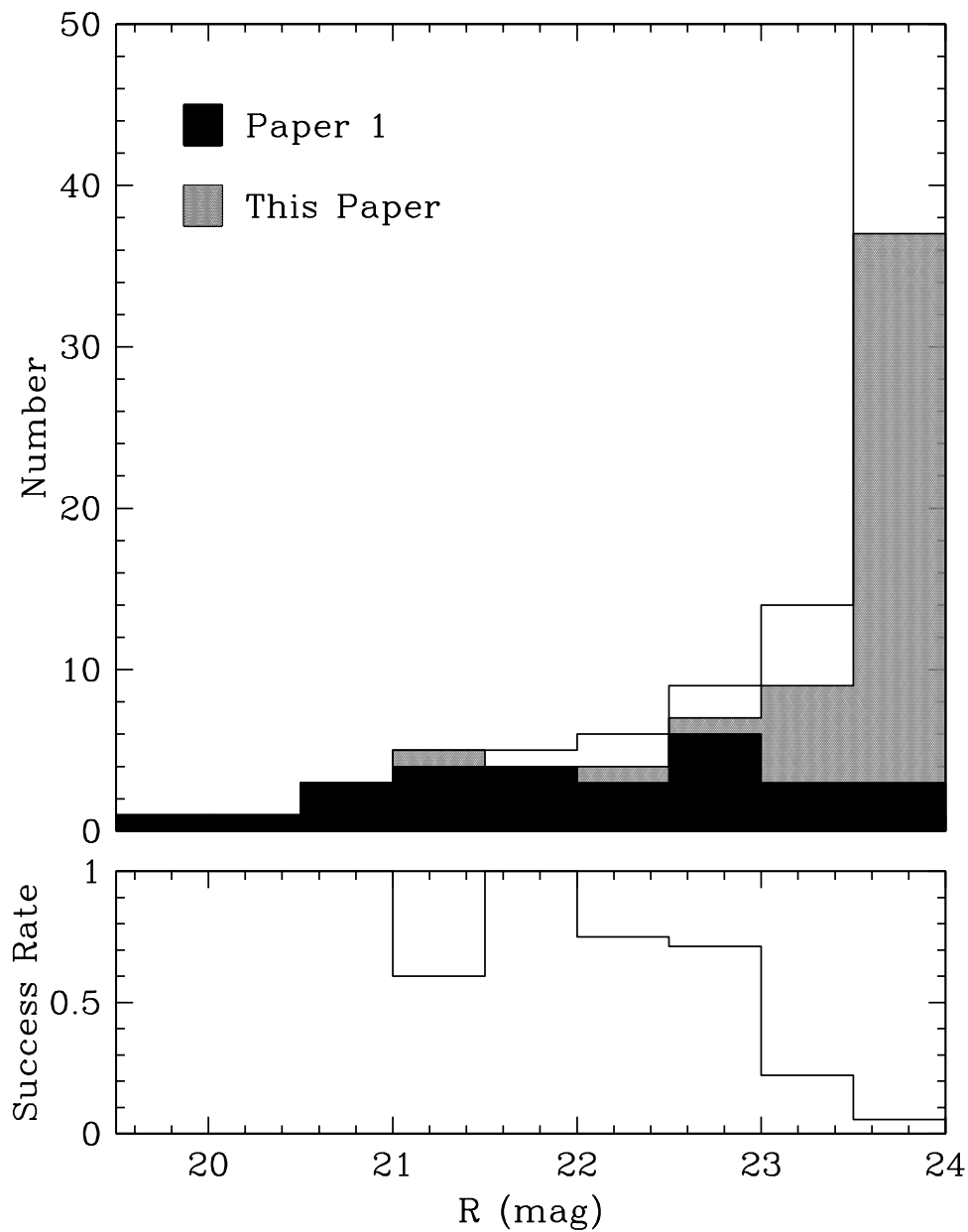}
\caption{
We have improved the completeness of our spectroscopic follow-up of our QSO candidates since Paper I.  {\em Top --} $R$-magnitude distribution of our quasar candidates.  Overplotted are objects with spectra from \citetalias{Glikman10} (solid) and this work (shaded, additional objects observed May 2010).  {\em Bottom --} Efficiency of the spectroscopic sample as a function of $R$-band magnitude. Note that our efficiency drops significantly for $R > 23$. }\label{fig:maghist}
\end{figure}

Even though we increased our spectroscopic completeness by more than a factor of two, only one $z=4.23$ QSO ($R=23.20$) was found in the NDWFS field. No QSOs were found in the DLS field.  In Figure \ref{fig:maghist} we show the $R$-magnitude distribution of our candidates. On top we show the distribution of candidates with spectra from \citetalias{Glikman10} in solid bins and the newly obtained spectra in shaded bins. The bottom panel shows the efficiency of finding QSOs as a function of magnitude.  It is clear that our selection efficiency drops significantly beyond $R=23$.  


The spectra that were not classified as QSOs fell into three categories:  (1) featureless, unidentifiable spectra, (2) Lyman Break Galaxies (LBGs) at similar redshifts ($z\sim 3-4$), and (3) carbon stars.  The objects with featureless spectra are not QSOs because they have smooth continua to the shortest wavelength end of the LRIS red camera ($\lambda \gtrsim5600$\AA) and no lines were identified.  The typical rms of our spectra is $\sim10^{-19}$ erg s$^{-1}$ cm$^{-2}$ \AA$^{-1}$, and the faintest \lya\ lines in our survey has a peak flux density of $\sim 4\times10^{-18}$ erg s$^{-1}$ cm$^{-2}$  \AA$^{-1}$.  With signal-to-noise ratio of $\gtrsim10$, Ly$\alpha$ has an unmistakeable signature in these spectra and its absence virtually guarantees that the object is not a QSO. 

LBGs at the same redshifts as the QSOs in our survey are a natural contaminant since their spectra also show the strong spectral break at $\sim 6000$\AA\ due to absorption by the \lya\ forest.  As we probe fainter fluxes we intersect the bright-end of the LBG luminosity function.  \citet{Steidel99}, using a somewhat different color selection at $z\sim 4$, estimated a surface density of $\sim 15$ LBGs per $\deg^2$ down to $I_{\rm AB} \simeq 23.5$ mag and $\sim 65$ LBGs per $\deg^2$ down to $I_{\rm AB} \simeq 24$.  This amounts to $56$ and $244$ LBGs expected in our area for the two magnitude limits, respectively, which is roughly consistent with the number of interlopers in our faintest bins.  

Carbon stars are common interlopers in high-redshift QSO surveys \citep{Richards02} and vice-versa \citep{Downes04}.  A more detailed discussion of our spectroscopic results will be presented in Paper III. (Glikman et al., in preparation).

\section{Results}

In \citetalias{Glikman10} we presented two versions of the QLF using absolute magnitudes computed from $K$-corrections as well as directly from the spectra.  Since the $K$-corrections rely on a QSO template, distribution of UV continuum slopes, and distribution of \lya\ equivalent widths, the latter two of which are modeled as Gaussian distributions, the luminosity derived from these $K$-corrections is heavily dependent on assumptions about the spectral properties of QSOs.  The measurement of $M_{1450}$ from spectrophotometry more directly reflects the true luminosity of each QSO.  Therefore, going forward we only use $M_{1450}$ measured from spectroscopy.


We extend our QLF to higher luminosities using QSOs from SDSS. In \citetalias{Glikman10} we compared our faint-end QLF with the SDSS-based QLF from \citet{Richards06} and \citet{Fontanot07}.  Although both bright-end QLFs used the same QSOs from SDSS, they derived volume densities that differed by more than a factor of 2 at their faintest bins.  This was due to different methods of deriving the selection function.   \citet{Fontanot07} used the QSO template from \citet{Cristiani90} and line and continuum properties based on the SDSS spectroscopic sample itself.  \citet{Richards06}, using a method more similar to ours, built a library of model QSO spectra with Gaussian distributions of line and continuum properties as we have done.  Therefore, to construct a QLF that will consistently span the full range of luminosities we use the selection function derived by \citet{Richards06} except that we re-compute $M_{1450}$ for the SDSS QSOs directly from their spectra.

\begin{figure}
\epsscale{1}
\plotone{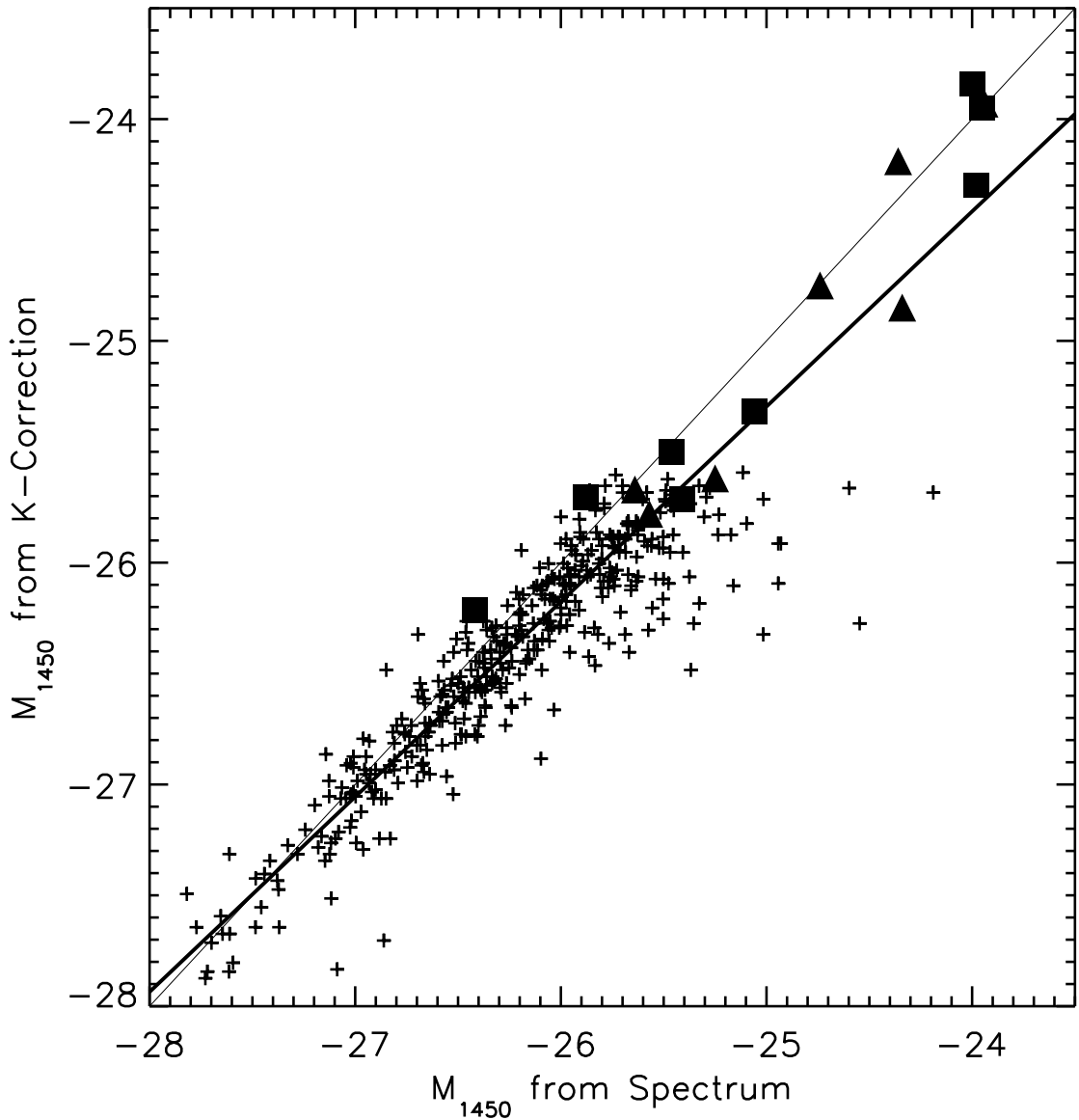}
\caption{The $K$-corrected absolute magnitudes at rest-frame 1450\AA\ versus the $M_{1450}$ measured directly from the QSO spectra.  The SDSS QSOs are represented with plus signs, squares are DLS QSOs, and triangles are QSOs from NDWFS. The thick line is the best fit relationship to all the points, with a slope of 0.9 and an offset of $-3.3$.  The thin line shows a one-to-one relation.  
$M_{1450}$ measured from spectrophotometry are significantly fainter (particularly at lower luminosities) than those derived from $K$-corrections.  This may be because the QSO templates used were derived from high-luminosity objects, yet are applied to the full range of source luminosities in our sample.}\label{fig:sdss_mag_compare}
\end{figure}

We identified QSOs from \citet[Table 5]{Richards06} with $3.8<z<5.2$ to match the redshift range of our QSO selection; 399 QSOs were selected.  We obtained their spectra from the SDSS database and measured $M_{1450}$ from each spectrum directly in the same manner as done for our faint QSOs.  Figure \ref{fig:sdss_mag_compare} compares $M_{1450}$ from $K$-corrected magnitudes \citep[using Equation 3 from][]{Richards06} with the spectrophotometrically derived values.  We include our faint QSOs as squares (DLS) and triangles (NDWFS).  We find that measuring $M_{1450}$ from $K$-corrections introduces a bias that overestimates the luminosities of QSOs.  The effect appears to be more pronounced at fainter luminosities.  This may be because the QSO template used to derive $K$-corrections is composed of luminous QSOs.  Luminous QSOs have more prominent blue bumps and therefore have stronger rest-frame UV emission than lower luminosity QSOs.  Fainter objects have weaker blue bumps and may be overcorrected by the $K$-correction technique, resulting in the bias seen in Figure \ref{fig:sdss_mag_compare}.

We use these new absolute magnitudes to compute $\Phi(M)$ using the value of the selection function from Table 5 of \citet{Richards06} and computing the available volume for each quasar, $V_a$, with the limiting magnitude of $i=20.2$.  We bin the QLF in one-magnitude bins and find that the faint end of the SDSS QLF matches quite well with the bright end of our survey (Figure \ref{fig:combined_qlf}).  Note that the faintest bin of the SDSS QLF is an outlier, likely due to incompleteness.  The open red circles show the QSO volume density derived in \citetalias{Glikman10}.  Table \ref{tab:lumfunc} lists the space densities of the binned QLF from SDSS and our QSOs.  We compute uncertainties for the bins with small numbers of QSOs ($<50$) using \citet{Gehrels86}. Our improved spectroscopic completeness lowers the space density of the lowest luminosity bin by a factor of 12, or approximately 1.2 sigma relative to the poorly constrained value presented in \citetalias{Glikman10}.   

\section{The Shape of the QLF}

We fit the QLF using the STY maximum likelihood method \citep{Efstathiou88}.  Initially we fit a single power-law function, which was the best-fit function found in \citetalias{Glikman10}.  The likelihood is maximized for a single power-law with $\alpha=-3.3\pm 0.1$ at 68\% confidence, $\pm 0.2$ at 90\% confidence, and $\pm 0.3$ at 99\% confidence.  This value is effectively a measure of the bright-end slope of the QLF and is consistent with \citet{Croom04} for the $0.4<z<2.1$ range.

Similarly, we measured the slope of a single power-law at the faint end, using only quasars fainter than $M_{1450} =-24$.  The likelihood is maximized with $\beta=-1.6\pm0.2$ at 68\% confidence,  $+0.3, -0.4$ at 90\% confidence, and $+0.4,-0.6$ at 99\% confidence.

\begin{figure}
\plotone{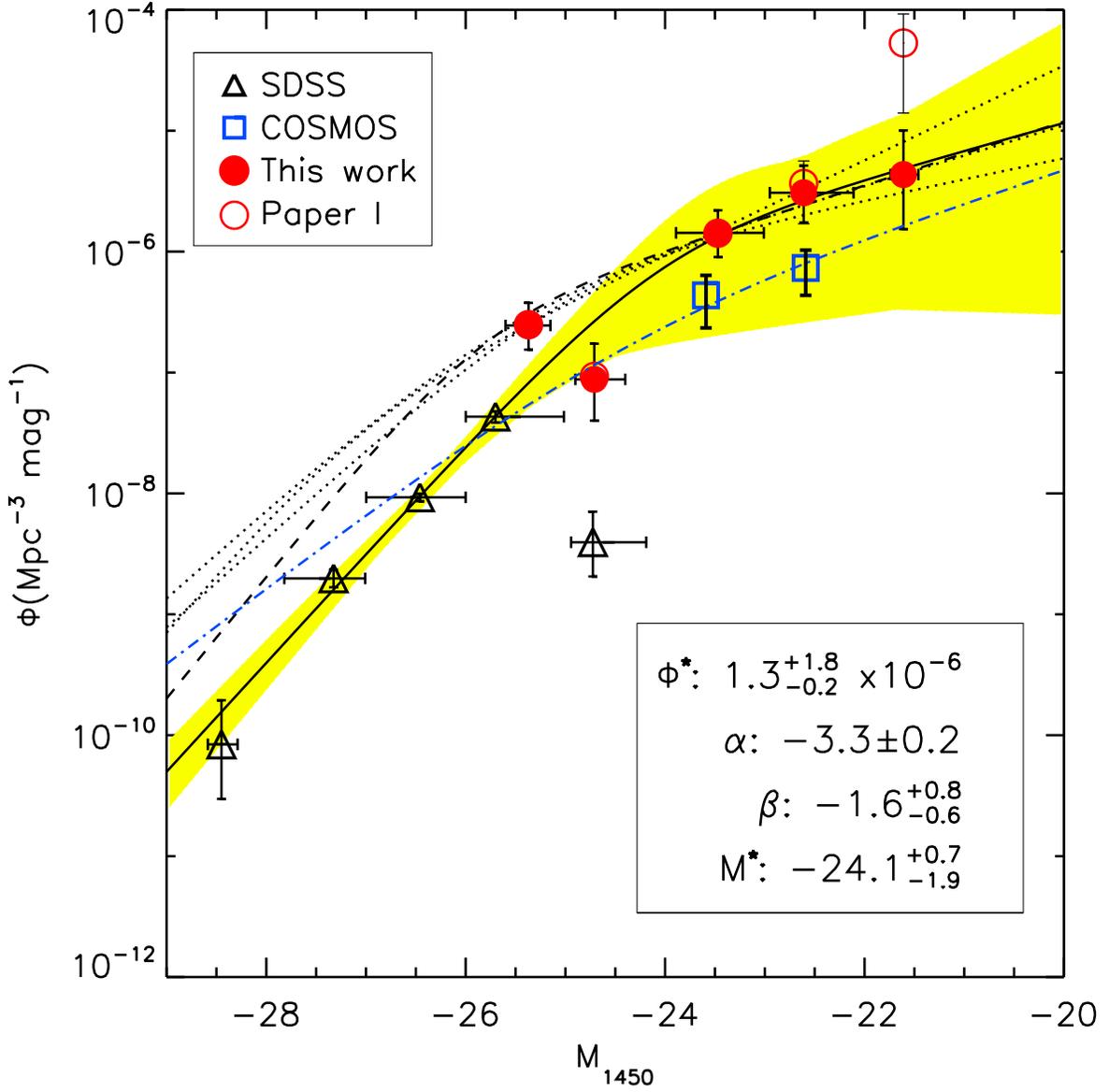}
\caption{The QSO luminosity function at $z\sim4$, combining the QLF from SDSS ({\em triangles}) and our new, updated data ({\em filled red circles}).  The points are plotted at the mean luminosity of each bin.  We overplot the QLF from Paper I  with {\em open red circles}, to show the change with increased spectroscopic completeness.  The {\em blue squares} are the space densities of $z\sim 4$ QSOs from \citet{Ikeda10} and the dot-dash line is their best-fit double power-law. The lower right-hand legend lists the best-fit parameters to a double power-law ({\em solid line}) shaded region represents the $1\sigma$ uncertainties.  Dashed and dotted lines show the $z\sim3$ QLF from \citet{Siana08}, representing the different fits to their QLF.  While the faint-end slopes agree with our measurements, our normalization, $\Phi^*$, is higher than \citet{Ikeda10} by a factor of $\sim 4$.}\label{fig:combined_qlf}
\end{figure}

\begin{figure}
\plotone{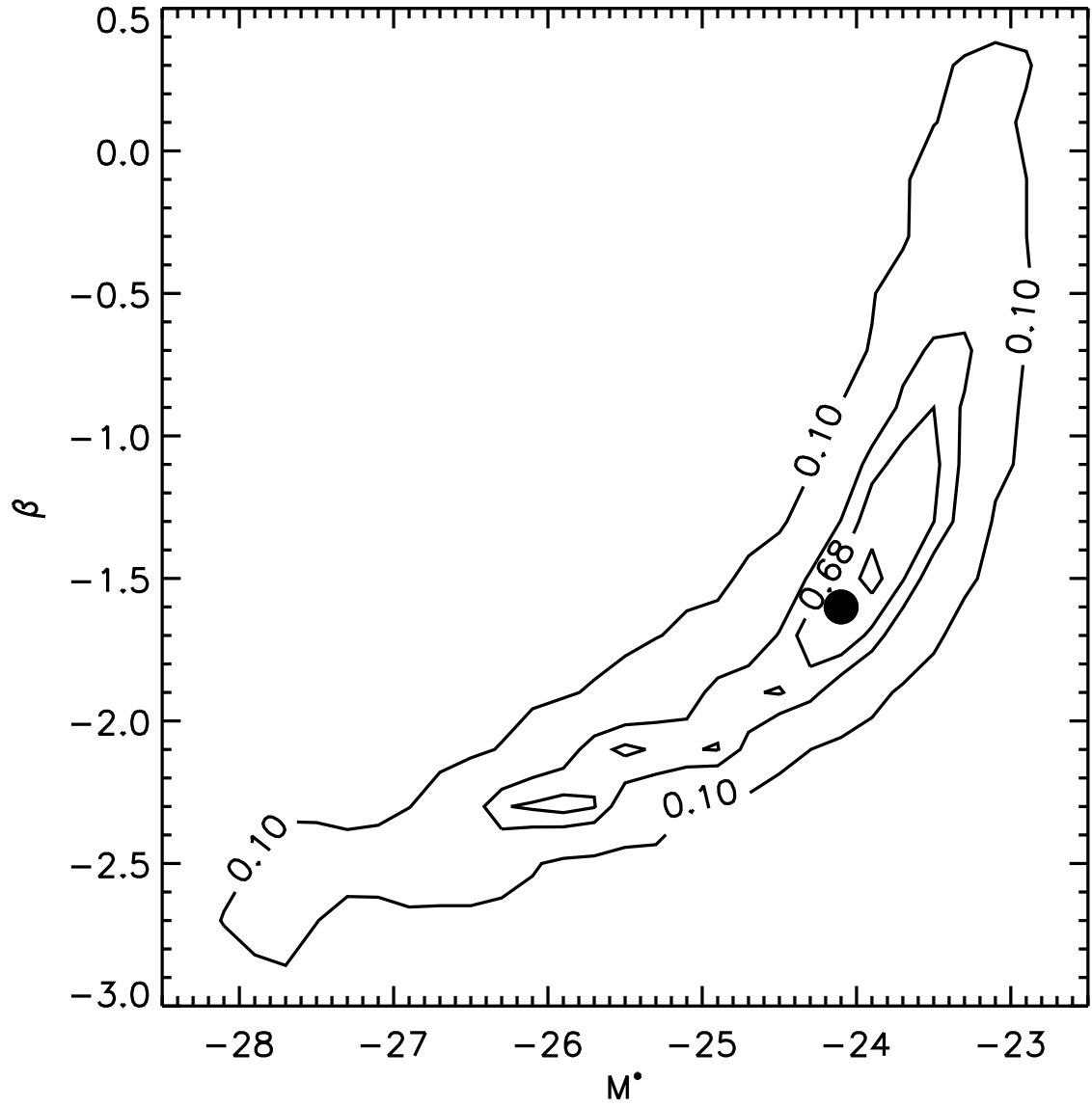}
\caption{We show the probability density of the faint-end slope, $\beta$, vs. break absolute magnitude, $M^*$, based on 10,000 Monte Carlo simulations of our binned QLF.  Contour levels are 10\%, 50\%, 68\% and 90\% of the peak density.  The best-fit parameters to the real QLF is plotted with a filled circle.}\label{fig:beta_mstar}
\end{figure}

We also fit the binned QLF to a double power-law function,
\begin{equation}
\Phi(M) = \frac{\Phi(M^*)}{10^{0.4(\alpha+1)(M-M^*)} + 10^{0.4(\beta+1)(M-M^*)}},
\end{equation}
omitting the outlying SDSS point at $M_{1450} = -24.5$.  The resultant curve is plotted in Figure \ref{fig:combined_qlf} with a solid black line.  The shaded region represents the $1\sigma$ uncertainties, which we determine by simulating 10,000 random QLFs based on our binned  $\Phi(M)$ measurements and their associated errors.  The best-fit parameters are: $\Phi^*= \Phi(M^*)=1.3^{+1.8}_{-0.2}\times10^{-6}$ Mpc$^{-3}$ mag$^{-1}$, $\alpha=-3.3\pm0.2$, $\beta=-1.6^{+0.8}_{-0.6}$, and $M^*=-24.1^{+0.7}_{-1.9}$ ($\chi^2 = 1.12$).  Figure \ref{fig:beta_mstar} shows the distribution of $\beta$ vs. $M^*$ from our simulations, with the best-fit value overplotted (filled circle).

The results from \citet{Ikeda10} are overplotted in Figure \ref{fig:combined_qlf} as blue squares.  The dot-dash line is their best-fit double power-law.  While our newly measured faint-end slope agrees with their value of $\beta=-1.67^{+0.11}_{-0.17}$, our normalization ($\Phi^*$) is higher by a factor of 4 and our break luminosity is fainter by 0.3 mag.  

We overplot in Figure \ref{fig:combined_qlf} the QLF at $z\sim 3$ from \citet{Siana08}, who selected QSOs using infrared and optical photometry in the SWIRE Legacy Survey \citep{Lonsdale03}.  The dashed line is their best-fit QLF to only the SWIRE data, while the dotted lines are fits to their SDSS+SWIRE QLF.  Both the bright- and faint-end slopes are comparable at $z\sim 3$ and $z\sim 4$.  The most striking evolution occurs at $M^*$ which, while poorly constrained, is $1-1.5$ magnitudes fainter at $z\sim 4$ than at $z\sim 3$, according to our fit.  

This is apparently at odds with the so-called ``cosmic downsizing'' where the space density of AGN peaks at different times for different luminosities. Low luminosity AGN are more abundant at low redshifts and the space density of high luminosity AGN peaks at earlier times.  This is seen in the evolution of the X-ray QLF \citep[e.g.,][]{Barger05} and in the optical at lower redshifts \citep[e.g.,][]{Croom09}.   However, this has not been well explored past the peak of the quasar epoch ($z\sim 2.5-3$).  Based on QSOs in SDSS Stripe82, two magnitudes deeper than \citet{Richards06}, \citet{Jiang06} did not see evidence for AGN downsizing at these redshifts.  
   
\section{The Contribution of Quasars to the UV Radiation Field at $z\sim 4$}

Following the same calculation as in \citetalias{Glikman10}, but using the new parameters of the double-power-law QLF, we calculate the emissivity of quasars at 1450\AA\ to be $\epsilon_{1450}=3.5^{+14.1}_{-2.4}\times10^{25} \ergs$ Hz$^{-1}$ Mpc$^{-3}$, $\sim 50\%$ of our earlier result.  \citet{Siana08} compute the specific luminosity density from QSOs at $z\sim3.2$ to be $\epsilon_{1450}=5.1\times10^{24}\ \ergs$ Hz$^{-1}$ Mpc$^{-3}$.  This is a factor of $\sim7$ lower than what we compute at $z\sim4$.

Following the formalism of \citet{Madau99}, $\dot{N}_{\rm IGM}=2.4\times10^{51}$ s$^{-1}$ ionizing photons are needed to ionize the IGM at $z=4.15$ (the median redshift of our survey).  Our parameterized QLF produces $\dot{N}_{\rm QSO} = 1.5^{+6.0}_{-1.0}\times10^{51}$ s$^{-1}$, or $\sim 60\pm40\%$ of the photons ionizing the IGM.  \citet{Siana08} argue that at $z\sim 3$ QSOs contribute just under half the needed photons to ionize the IGM, which is roughly consistent with our result.  In addition, \citet{Vanzella10} find that $z\sim4$ LBGs in the the Great Observatories Origins Deep Surveys \citep[GOODS;][]{Giavalisco04} account for $<20\%$ of the photons needed to ionize the IGM, leaving QSOs as the likely dominant ionizing source at this redshift.

\acknowledgments
We thank Gordon Richards for useful discussions on combining the SDSS and our QLFs and Meg Urry for helpful comments.  We are grateful to the staff of W. M. Keck observatory for their assistance during our observing runs. This work was supported in part by the NSF grants AST-0407448 and AST-0909182, and by the Ajax foundation.  The work of DS was carried out at Jet Propulsion Laboratory, California Institute of Technology, under a contract with NASA.  The research activities of AD and BTJ are supported by the NSF through its funding of the NOAO, which is operated by the Association of Universities for Research in Astronomy, Inc. under a cooperative agreement with the NSF.  This work makes use of image data from the NDWFS and the DLS as distributed by the NOAO Science Archive. KSL gratefully acknowledges the generous support of Gilbert and Jaylee Mead for their namesake fellowship.


\pagebreak

\begin{deluxetable}{cccr}
\tablecolumns{4}
\tablewidth{0pt} 

\tablecaption{Luminosity Function \label{tab:lumfunc}}

\tablenum{1}

\tablehead{\colhead{$M_{1450}$ Bin Center} & \colhead{$\langle M_{1450} \rangle$\tablenotemark{a}} & \colhead{$\Phi$} & \colhead{$N_{\rm QSO}$} \\ 
\colhead{(mag)} & \colhead{(mag)} & \colhead{($10^{-8}$ Mpc$^{-3}$ mag$^{-1}$)} & \colhead{} } 

\startdata
\cutinhead{SDSS}
$-$28.5 & $-$28.45 & $0.008 ^{+0.011}_{-0.005}$ & 2 \\
$-$27.5 & $-$27.33 & $0.20 ^{+0.04}_{-0.03}$ & 41 \\
$-$26.5 & $-$26.46 & $0.93 \pm0.07$ & 169 \\
$-$25.5 & $-$25.70 & $4.3 \pm0.5$ & 102 \\
$-$24.5 & $-$24.72 & $0.4 ^{+0.3}_{-0.2}$ & 4 \\
\cutinhead{NDWFS+DLS}
$-$25.5 & $-$25.37 & $24^{+13}_{-9}$ &  7\\
$-$24.5 & $-$24.71 & $8.8^{+8.5}_{-4.8}$&  3\\
$-$23.5 & $-$23.47 & $143^{+77}_{-53}$&  7\\
$-$22.5 & $-$22.61 & $307^{+208}_{-133}$ &  5\\
$-$21.5 & $-$21.61 & $434^{+572}_{-280}$ &  2\\

\enddata
\tablenotetext{a}{Mean magnitude of the QSOs in each bin.}
\end{deluxetable}

\end{document}